\documentclass[prl,notitlepage,twocolumn,longbibliography]{revtex4-2}
\usepackage{amsmath,amsthm,amssymb,amsfonts,graphicx}
\usepackage{tikz} 
\usetikzlibrary{positioning}
\usepackage{xcolor}
\usepackage{cancel}
\usepackage{braket}
\usepackage{calc}
\usepackage{bbold}

\begin{document}

\title{Beating Carnot efficiency with periodically driven chiral conductors}

\author{Sungguen Ryu}
\author{Rosa L\'{o}pez}
\author{Lloren\c{c} Serra}
\author{David S\'{a}nchez}
\affiliation{Instituto de F\'{i}sica Interdisciplinar y Sistemas Complejos IFISC (CSIC-UIB), E-07122 Palma, Spain \\
  \textnormal{Correspondence and requests for materials should be addressed to S.R.} (email: sungguen@ifisc.uib-csic.es)}

\date{\today}

\begin{abstract}
  \textbf{Abstract.} Classically, the power generated by an ideal thermal machine cannot be larger than the Carnot limit. This profound result is rooted in the second law of thermodynamics. A hot question is whether this bound is still valid for microengines operating far from equilibrium. Here, we demonstrate that a quantum chiral conductor driven by AC voltage can indeed work with efficiencies much larger than the Carnot bound. The system also extracts work from common temperature baths, violating Kelvin-Planck statement. Nonetheless, with the proper definition, entropy production is always positive and the second law is preserved. The crucial ingredients to obtain efficiencies beyond the Carnot limit are: i) irreversible entropy production by the photoassisted excitation processes due to the AC field and ii) absence of power injection thanks to chirality. Our results are relevant in view of recent developments that use small conductors to test the fundamental limits of thermodynamic engines.
\end{abstract}

\maketitle

\textbf{Introduction}

Quantum thermodynamics is a thriving field that is being nourished by the cross-fertilization of statistical mechanics, quantum information and quantum transport~\cite{kos13,QTbook1,QTbook2}. Interestingly enough, one of its major subjects of enquiry is still connected with the problem that Sadi Carnot analyzed two centuries ago~\cite{Carnot1824}, namely, what is the maximum amount of useful work achieved by a generic heat engine? The crucial difference is that the working substance in classical thermodynamic schemes is now a quantum system~\cite{benenti2017fundamental}. This challenges the paradigms of thermodynamics, which are in principle of universal validity, and calls for a revision of our definitions of heat, work and entropy~\cite{lud14,esp15,car16,bruch2016quantum,ludovico2016adiabatic,bruch2018landauer,Bhandari}.

In general, the efficiency of thermal machines at both macro or nano scales is limited by the entropy production that the second law of thermodynamics dictates to be a positive quantity. 
In classical thermodynamics the Clausius inequality for entropy production implies a maximum efficiency, the Carnot efficiency. 
However, this upper bound can in principle be surpassed  if we assume quantum coherence is also a resource for entropy production \cite{Scully862,PhysRevLett.119.170602}. The same argument also implies that a quantum heat engine can violate the Kelvin-Planck statement that no useful work can be extracted from common temperature reservoirs. 
Hence, understanding how the entropy resource can be controlled in different scenarios is key to achieve enhanced efficiencies in quantum coherent engines and refrigerators.

The widespread interest in quantum thermal machines is also grounded on the plethora of platforms where dynamics can be controlled at a microscopic scale, such as trapped ions~\cite{PhysRevLett.109.203006}, quantum dots~\cite{Kennes_2013}, single electron boxes~\cite{Koski13786}, optomechanical oscillators~\cite{PhysRevLett.112.150602}, QED circuits~\cite{PhysRevLett.112.076803} and multiterminal conductors~\cite{PhysRevB.83.085428,Entin2010Three}. A shared property of these setups is that the quantum system couples to external baths with which exchanges particles, energy or different degrees of freedom~\cite{Alicki_1979,Ronnie1984,Campisi_2015}. The baths have thus far been treated with well defined chemical potentials and temperatures. By contrast, the case of baths that are driven out of equilibrium are quite scarce~\cite{sanchez2019nonequilibrium,PhysRevLett.126.080603,Holubec}. Such investigation is desirable and timely, as the control and measurement of nanoscale systems driven out of equilibrium by high-frequency AC potential is being experimentally realized, providing single-electron sources. 
Specifically, we refer to the cases of so-called Levitons~\cite{keeling2006minimal} in fermionic quantum optics~\cite{Leviton2013,jullien2014quantum}, quantum-dot pumps for metrology~\cite{yamahata2019picosecond}, and flying qubits for quantum information processing~\cite{bauerle2018coherent}. The study about heat and energy currents carried by such single-electron sources is of recent interest~\cite{dashti2019minimal,vannucci2017minimal}.

\begin{figure}[t]
 \centering
 \includegraphics[width=\columnwidth]{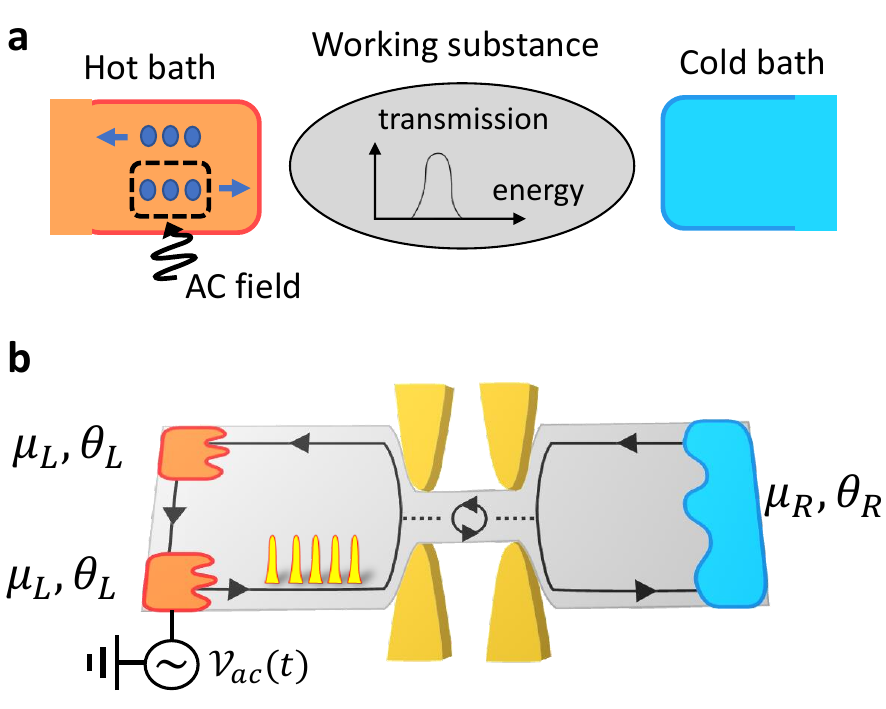}
 \caption{\textbf{Setup.} (a) Schematic of periodically driven chiral engine. In the hot bath, an AC field is selectively applied only to the electrons directed towards the working substance. The working substance is composed of a scatterer whose transmission probability has an arbitrary energy dependence.
     (b) An implementation using a chiral conductor with a localized energy level. A time-dependent AC voltage $\mathcal{V}_{\text{ac}}(t)$ is applied only to the lower left reservoir, hence realizing the selective AC driving. 
   Edge channels from left and right reservoir are tunnel coupled to the localized state.   
 The hot left reservoirs (cold right reservoir) have temperature $\theta_{L(R)}$ and chemical potential $\mu_{L(R)}$. }
 \label{fig:setup}
\end{figure}

In this work, we show that a generic class of periodicallty driven quantum devices can reach thermodynamic efficiencies that surpass the Carnot limit.
Our pump engine [see Fig.~\ref{fig:setup}(a)] consists of a scatterer of arbitrary energy-dependent transmission tunnel coupled to electronic hot and cold reservoirs in the presence of an external AC bias voltage.
An AC driving typically generates a finite input power that diminishes the efficiency. 

Our key idea to overcome this difficulty is to selectively apply an AC external field to the electrons depending on the direction, which can be implemented using a chiral conductor such as those created with topological matter~\cite{Kane2010Topological} [see Fig.~\ref{fig:setup}(b)].
This completely  avoids any AC input power, allowing a high efficiency of the quantum engine, in contrast to nonchiral cases.
Our main finding is that such photonic and chiral engine boosts the 
heat-machine efficiency above the Carnot bound 
due to a remarkable interplay between nonequilibriumness and chirality. Under these circumstances caution is needed in the thermodynamical description that has to be adapted for  a nonthermal bath. 
We adopt the Floquet scattering matrix approach~\cite{moskalets2002floquet,moskalets2011scattering} for electric and heat currents
and also need a generalized definition of entropy production
based on Shannon formula for the incoming and outgoing electron 
distributions in each terminal \cite{bruch2018landauer}.

\vspace{2mm}
\textbf{Results}

{\it Floquet scattering matrix---}
The setup in Fig.~\ref{fig:setup}(b) consists of a localized energy level (a quantum dot or impurity) tunnel linked to chiral edge conducting modes with a tunnel rate $\Gamma$.
The chiral modes are connected to three reservoirs where the left two reservoirs play the role of the hot reservoir in Fig.~\ref{fig:setup}(a) with the selective AC driving. This selectivity is realized by applying an AC gate voltage only to the lower reservoir, exploiting the chirality of the conductor.
  The electrons deep in the left upper or left lower reservoir are described with the Fermi-distribution $f_L(\mathcal{E})$ and the electrons deep in the right reservoir are described with $f_R(\mathcal{E})$, where $f_\beta(\mathcal{E})=1/[1+\exp\{(\mathcal{E}-\mu_\beta)/(k_B \theta_\beta)\}]$ for $\beta =L, R$,  $\theta_{\beta}$ and $\mu_\beta$ are temperature and chemical potential of the reservoir $\beta$, and $k_B$ is Boltzmann constant.
We assume $\theta_L\ge \theta_R$.
Below, we present the scattering formalism in a two-terminal form, treating the left two reservoirs as a single bath, as we are only interested in the net charge (or heat) currents flowing into the terminals.

In the static case, the scatterer has reflection ($r_{\text{st}}$) and transmission ($t_{\text{st}}$) amplitudes that depend on energy $\mathcal{E}$: 
 \begin{equation}
  \mathcal{S}^{\text{(st)}}(\mathcal{E})= \begin{bmatrix}
     r_{\text{st}}(\mathcal{E}) & t_{\text{st}}'(\mathcal{E}) \\
     t_{\text{st}}(\mathcal{E}) & r_{\text{st}}'(\mathcal{E})
   \end{bmatrix},
 \end{equation}
where the scattering amplitudes are denoted with $'$ when input is from the right reservoir. 

In the presence of the external AC bias voltage $\mathcal{V}_{\text{ac}}(t)$, whose time average is 0 and frequency is $\Omega/(2\pi)$, an electron of energy $\mathcal{E}$ in the left input reservoir first gains or loses kinetic energy by absorbing ($n>0$) or emitting ($n<0$) $|n|$ photons with the transition amplitude $a_n$~\cite{dubois2013integer,kohler2005driven,platero2004photon},
\begin{equation}
  \label{eq:an}
  a_n = \frac{\Omega}{2\pi} \int_0^{2\pi/\Omega} dt \, e^{i \phi_{\text{ac}}(t) } e^{i n \Omega t}\;.
\end{equation}
(See Supplementary Information (SI) for a derivation.)
$n=0$ describes the direct process in which photons are neither absorbed nor emitted.  $\phi_{\text{ac}}(t) \equiv -\int_{-\infty}^t dt' e\mathcal{V}_{\text{ac}}(t')/\hbar $ is the phase due to the AC voltage. 
The Floquet scattering matrix~\cite{moskalets2002floquet,moskalets2011scattering} $\mathcal{S}_{\alpha \beta} (\mathcal{E}_n , \mathcal{E})$ describes the whole process whereby an electron of energy $\mathcal{E}$ from reservoir $\beta$ absorbs/emits $|n|$ photons, scatters off the localized level, and finally enters reservoir $\alpha$ with energy $\mathcal{E}_n\equiv \mathcal{E}+ n \hbar\Omega$,
\begin{equation}
\begin{array}{ll}
 \mathcal{S}_{RL}(\mathcal{E}_n, \mathcal{E})
 = a_n\, t_{\text{st}}(\mathcal{E}_n)\,, &
 \mathcal{S}_{LL} (\mathcal{E}_n, \mathcal{E})
 =a_n\, r_{\text{st}}(\mathcal{E}_n)\,,  \\
\rule{0cm}{0.4cm}
 \mathcal{S}_{LR}(\mathcal{E}_n, \mathcal{E})  = \delta_{n,0}\, t'_{\text{st}}(\mathcal{E})\,,  &
\mathcal{S}_{RR}(\mathcal{E}_n, \mathcal{E}) = \delta_{n,0}\, r'_{\text{st}}(\mathcal{E})\,. 
\end{array}
\label{eq:SF}                     
\end{equation}
These satisfy unitarity of Floquet scattering matrix, $\sum_{n \alpha} |\mathcal{S}_{\alpha \beta}(\mathcal{E}_n, \mathcal{E})|^2  =1$, and $\sum_{n \beta} |\mathcal{S}_{\alpha \beta}(\mathcal{E}, \mathcal{E}_{-n})|^2  =1$ (see SI).

\begin{figure*}[t]
  \centering
  \includegraphics[width=0.95\textwidth]{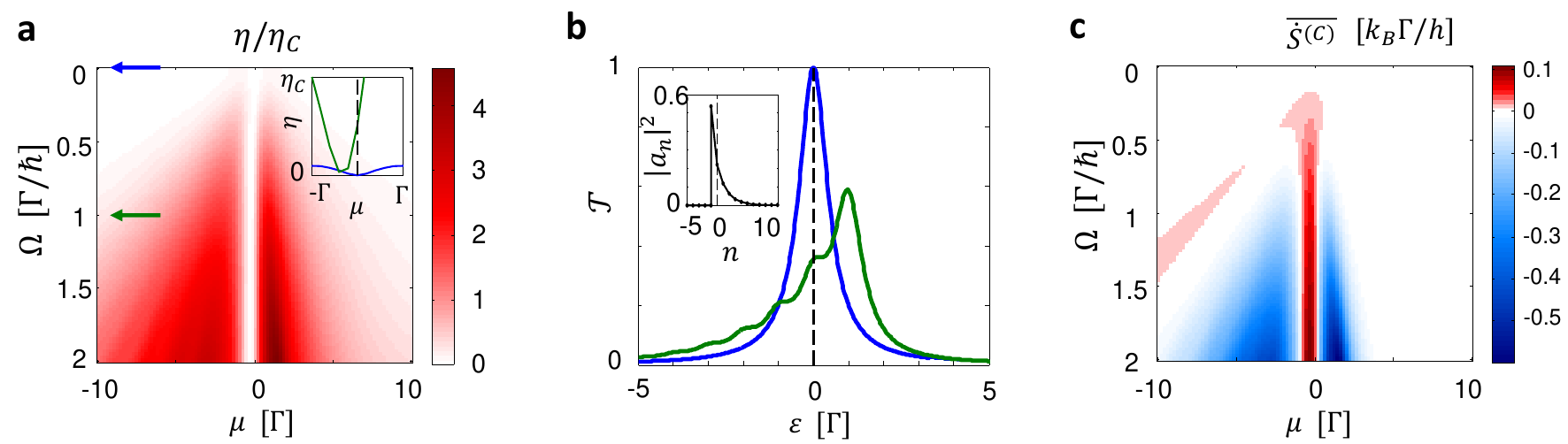}
  \caption{\textbf{Efficiency enhancement beyond the Carnot limit by the AC voltage driving.} Here AC voltage is chosen as periodic Lorentzian pulses which generate Levitons. 
    (a) Efficiency when tuning the driving frequency $\Omega$ and the average chemical potential $\mu\equiv (\mu_L+\mu_R)/2$ (measured from the resonance level). Inset: Plot near $\mu=0$ (dashed line) for $\Omega=0$ (blue) and $\Omega = \Gamma/\hbar$ (green) as indicated with arrows in the main panel.    
    (b) Total photoassisted transmission probability $\mathcal{T}$ for an electron of energy $\mathcal{E}$ (measured from the resonance level) incoming from the left reservoir for $\Omega=0$ (blue) and $\Omega = \Gamma/\hbar$ (green).
    Inset: the photoassisted transition probabilities $|a_n|^2$.
    (c) The entropy production rate assuming Clausius relation, $\overline{\dot{S}^{(C)}}$.
    Here, $(\theta_L+\theta_R)/2 = 0.25\Gamma/k_B$, $\theta_L-\theta_R = 0.1\theta$, $w= 0.05\times 2\pi/\Omega$, and the DC chemical potential bias is chosen for the maximal power generation.
    }
  \label{fig:eng-chiral}
\end{figure*}

For the results below, it is useful to define the mean number of photons absorbed during photoassisted scattering process,
$\braket{n (\mathcal{E})}_{\alpha \beta} = \sum_n n |\mathcal{S}_{\alpha \beta}(\mathcal{E}_n, \mathcal{E})|^2$.
The total transmission probability from left [right] reservoir, including  all photoassisted transitions, becomes
$\mathcal{T}(\mathcal{E}) \equiv \sum_n |\mathcal{S}_{RL} (\mathcal{E}_n, \mathcal{E})|^2$
[$\mathcal{T}'(\mathcal{E}) \equiv \sum_n |\mathcal{S}_{LR} (\mathcal{E}_n, \mathcal{E})|^2$].

{\it Charge and heat flows and efficiency---}
The time-averaged charge, heat and energy currents flowing into the $\alpha$ reservoir are determined by the above mentioned Floquet scattering matrix as~\cite{moskalets2002floquet}
\begin{align}
  \overline{I_e^\alpha}
  &= e \int_0^\infty \frac{d \mathcal{E}}{h} \,  \sum_{\beta,n}\big\{- \delta_{\alpha \beta}\delta_{n0} + |\mathcal{S}_{\alpha\beta}(\mathcal{E}_n, \mathcal{E})|^2\big\} f_\beta(\mathcal{E} ) , \label{eq:I-e-alpha}\\
  \overline{I_h^\alpha}
  &= \int_0^\infty \hspace{-0.1cm} \frac{d \mathcal{E}}{h}  \sum_{\beta,n}  \big\{ -\delta_{\alpha\beta}\delta_{n0} \nonumber \\
  & \qquad \qquad \qquad
  + |\mathcal{S}_{\alpha\beta}(\mathcal{E}_n, \mathcal{E})|^2 \big\} (\mathcal{E}_n-\mu_\alpha) f_ \beta(\mathcal{E}) , \label{eq:I-h-alpha}\\
  \overline{I^\alpha_u}
  &= \int_0^\infty \frac{d \mathcal{E}}{h} \, \sum_{\beta, n} \{- \delta_{\alpha \beta} \delta_{n0} + |S_{\alpha \beta}(\mathcal{E}_n , \mathcal{E})|^2\} \mathcal{E}_n f_\beta(\mathcal{E}) ,
    \label{eq:I-u-alpha}
\end{align}
where $e \,(<0)$ is the electron charge and $h$ is the Planck constant.
Using unitarity, we obtain 
\begin{equation}
     \label{eq:I-e} 
 \overline{I_e^R} = \frac{e}{h} \int_0^\infty d \mathcal{E} \, \big\{
  \mathcal{T}(\mathcal{E}) f_{L}(\mathcal{E})
                     - \mathcal{T}'(\mathcal{E}) f_R(\mathcal{E})\big\}.
\end{equation}
Current conservation, $\overline{I_e^R} +\overline{I_e^L} =0$, is satisfied over one driving period since no charge piles up in the steady state.
 Expanding $\mathcal{E}_n-\mu_{\alpha}$ into $\mathcal{E}-\mu_\alpha$ and $n \hbar\Omega$, and using again unitarity, we obtain
\begin{align}
 \overline{I_h^R} &= \frac{1}{h} \int_0^\infty d \mathcal{E} \, (\mathcal{E} -\mu_R)
    \big\{\mathcal{T}(\mathcal{E}) f_L(\mathcal{E})    -\mathcal{T}' (\mathcal{E})f_R(\mathcal{E}) \big\}   \nonumber \\
  &\quad +  \frac{\Omega}{2\pi} \sum_{\beta=L,R} \int_0^\infty d \mathcal{E} \,
     \braket{n(\mathcal{E})}_{R\beta} f_\beta(\mathcal{E})\;.
       \label{eq:I-h} 
\end{align}
$\overline{I_h^L}$ is determined by Eq.~(\ref{eq:I-h}) with substitutions $(\mathcal{E} -\mu_R) \rightarrow -(\mathcal{E} -\mu_L)$ and $\braket{n(\mathcal{E})}_{R\beta} \rightarrow \braket{n(\mathcal{E})}_{L\beta}$.
Note that heat [charge] current is positive when the electrons flow into [out of] the reservoir.

We compute the power associated to DC and AC voltage bias, in order to obtain the net generated power by the chiral 
conductor $\overline{P_{\text{gen}}}$ over a period. The DC voltage bias applied against the current flow generates electrical power $\overline{P_e} = - (\mu_L-\mu_R) \overline{I_e^R}/e$~\cite{benenti2017fundamental}.
The time-averaged power injected into the conductor by the AC voltage is related to the energy currents 
as $\overline{P_{\text{in}}} = \overline{I^L_u} +\overline{I^R_u}$ due to energy conservation during one AC period~\cite{dare2016time,ludovico2016dynamics}.
The net generated power is determined by the DC power subtracted by the injected AC power,
\begin{equation}
  \overline{P_{\text{gen}}}
  =\overline{P_e} -\overline{P_{\text{in}}},   \label{eq:Pgen} 
\end{equation}
Since the DC power $\overline{P_e}$ is dictated by the electrical flow and the injected AC power by the energy currents, the first law of thermodynamics demands that the net generated power is given by the heat fluxes as  $\overline{P_{\text{gen}}}=-(\overline{I_h^R}+\overline{I_h^L})$.
Using Eqs.~(\ref{eq:I-e})--(\ref{eq:Pgen}),  the input power is written in terms of Floquet scattering matrix as
\begin{equation}
 \label{eq:P-in}
  \overline{P_{\text{in}}}
  = \sum_{\alpha \beta}\int \frac{d \mathcal{E}}{h}
  \hbar\Omega \braket{n(\mathcal{E})}_{\alpha \beta} 
  f_\beta(\mathcal{E}).
\end{equation}
This clarifies the time-averaged power associated to AC voltage in a general Floquet scattering situation.
The term $\hbar \Omega \braket{n(\mathcal{E})}_{\alpha \beta}$ describes the energy change of the electron incoming from reservoir $\beta$ and outgoing to $\alpha$ via photon absorptions/emissions. The factor $(d \mathcal{E}/h) f_\beta (\mathcal{E})$ accounts for the injection rate of electron in the energy window $[\mathcal{E}, \mathcal{E}+d \mathcal{E}]$ from reservoir $\beta$.

A salient feature of our chiral setup is that the power associated to AC voltage is zero, $\overline{P_{\text{in}}}=0$, regardless of the form of the AC driving $\mathcal{V}_{\text{ac}}(t)$ and the scatterer. This is a consequence of the fact that the mean photon number $\braket{n} \equiv \sum_n n |a_n|^2$ involved in the photon absorption/emission by AC voltage is zero,
$\braket{n} = 0$ (because $\braket{n} = e\overline{\mathcal{V}_{\text{ac}}}/(\hbar\Omega)=0$ as derived in SI) and the Floquet scattering matrix has the form of Eq.~(\ref{eq:SF}) for a chiral system.
Then, the mean number of absorbed photons by an electron  traveling from the left reservoir is also zero, as  $ \braket{n (\mathcal{E})}_{LL}+\braket{n (\mathcal{E})}_{RL} = \sum_n n |a_n|^2 [|t_{\text{st}}(\mathcal{E}_n)|^2 + |r_{\text{st}}(\mathcal{E}_n)|^2] = \braket{n}$. Besides, an electron from the right reservoir does not absorb nor emit photons, thus $\braket{n(\mathcal{E})}_{LR}=\braket{n(\mathcal{E})}_{RR}=0$.
Hence, $\overline{P_{\text{in}}}=0$ according to Eq.~(\ref{eq:P-in}).
Contrarily, in the nonchiral case (see SI for details), $\overline{P_{\text{in}}}$ is generally positive. In fact, in the limit of slow driving ($\Omega \ll \Gamma/\hbar$) and zero temperature case, the power injection is in the form of Joule's law, $\overline{P_{\text{in}}}=  |t_{\text{st}}(\mu_L)|^2 \frac{e^2}{h} \overline{\mathcal{V}_{\text{ac}}^2(t)}$, hence positive.
The positive $\overline{P_{\text{in}}}$ diminishes the net generated work, and thus efficiency in nonchiral conductors.

When our system satisfies $\overline{I_h^L} <0$ and  $\overline{P_{\text{gen}}}>0$, it works as a heat engine with
efficiency $\eta$~\cite{dare2016time},
\begin{equation}
  \label{eq:eta}
    \eta = \frac{\overline{P_{\text{gen}}}}{-\overline{I_h^L}}\;.
\end{equation}

In static situations, it is common to compare $\eta$ with the Carnot efficiency $\eta_C= 1-\theta_R/\theta_L$, i.e., the upper bound dictated by the positivity of the entropy production from Clausius relation,
\begin{equation}\label{eq:clausius}   \overline{\dot{S}^{(C)}}=\sum_\alpha \frac{\overline{I^\alpha_h}}{\theta_\alpha}\;.
\end{equation}
However, caution is needed for such comparison in nonequilibrium situations. The fact of having nonthermal contacts prevents us from considering the entropy production written as in Eq.~(\ref{eq:clausius}). Indeed, our results indicate for some regime of parameters $ \overline{\dot{S}^{(C)}}<0$ (see Fig.~\ref{fig:eng-chiral}(c)), seemingly violating the second law of thermodynamics. To get a deeper insight, we employ the entropy production defined using Shannon entropy of the unperturbed (incoming) distribution function $f_\alpha (\mathcal{E})$ and the AC driven nonequilibrium distribution function 
$f^{(\text{out})}_\alpha (\mathcal{E})  
= \sum_{n,\beta} |S_{\alpha \beta}(\mathcal{E}, \mathcal{E}_{n})|^2   f_\beta (\mathcal{E}_{n})$,
recently proposed in Ref.~\cite{bruch2018landauer} for a periodically driven system,
  \begin{equation}
  \overline{\dot{S}}
  = \frac{k_B}{h} \sum_{\alpha=L,R} \int d \mathcal{E}
  \left(- \sigma [ f_\alpha (\mathcal{E})]
  + \sigma [ f_\alpha^{(\text{out})} (\mathcal{E})] \right).
    \label{eq:dS-f} 
\end{equation}
Here $\sigma[f] \equiv -f \ln f - (1-f)\ln (1-f)$ is the binary Shannon entropy function measured in nats.

A further step clarifies the deviation $ \delta \overline{\dot{S}} \equiv \overline{\dot{S}} -\overline{\dot{S}^{(C)}}$ of Eq.~(\ref{eq:dS-f}) from its counterpart deduced from the Clausius relation Eq.~(\ref{eq:clausius}).
Remarkably, we find a simple relation between the deviation and the photon number uncertainty $ \delta n \equiv  \sqrt{\sum_n (n-\braket{n})^2 |a_n|^2}$, in the regime of small biases
$k_B|\theta_L -\theta_R|, \, |\mu_L-\mu_R| \ll k_B \theta_L$
and small energy uncertainty induced by the AC voltage
$ \delta n \, \hbar \Omega \ll k_B \theta_L $,
(see SI for the details)
\begin{equation}
  \delta \overline{\dot{S}}
  \approx \frac{(\delta n\, \hbar \Omega)^2}{2 h \theta_L}.
    \label{eq:ddS}
\end{equation}
The energy uncertainty $\delta n\, \hbar\Omega$ is equal to the root mean square of the AC potential energy, $e(\overline{\mathcal{V}_{\text{ac}}^2})^{1/2}$.
Hence, Eq.~(\ref{eq:ddS}) quantifies how the initially thermal electrons in the left reservoir are driven into the nonthermal state due to energy uncertainty induced by the AC voltage.

\begin{figure}[t]
  \centering
  \includegraphics[width=0.95\columnwidth]{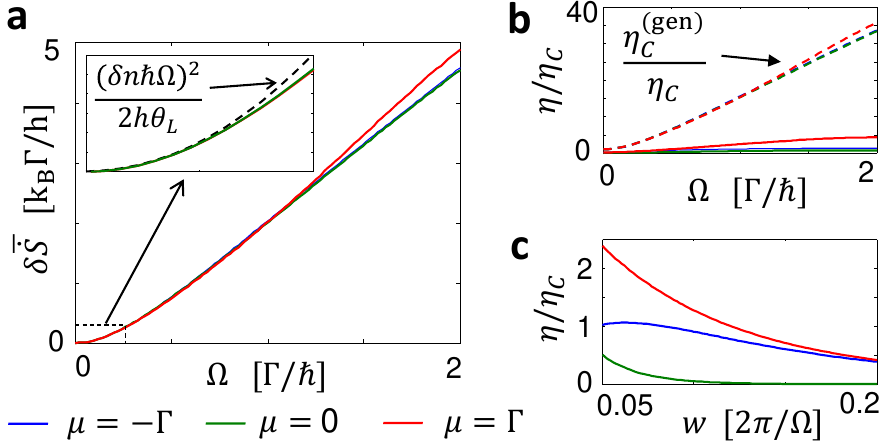}
  \caption{\textbf{Departure from Clausius relation.} (a) Deviation of the entropy production $\delta \overline{\dot{S}}$ from Clausius relation, in the situation of Fig.~\ref{fig:eng-chiral}. Inset: Comparison with the slow-driving limit, Eq.~(\ref{eq:ddS}) (dashed line) for $\Omega \in [0, k_B\theta_L/\hbar]$. (b) Efficiency in comparison with the upper bound determined by positivity of entropy production, $\eta_C^{\text{(gen)}} = \eta_C + \theta_R\delta \overline{\dot{S}}/|\overline{I_h^L}|$ (dashed lines). (c) Efficiency when tuning the temporal width $w$ of the Leviton, while fixing $\Omega=\Gamma/\hbar$. 
}
  \label{fig:ddS}
\end{figure}

{\it Numerical Results---}
Fig.~\ref{fig:eng-chiral} shows the result of our numerical calculations.
In the static situation, the scattering matrix $\mathcal{S}^{\text{(st)}}$ is modeled as a Breit-Wigner resonance whose full width at half maximum is $\Gamma$.
As an illustration showing most dramatic effects, e.g. dynamical electron-hole symmetry breaking as shown below, of the AC driving, we consider the AC voltage bias of periodic Lorenztian pulses correspondig to the protocol generating Levitons of charge $e$~\cite{keeling2006minimal,dubois2013integer}, 
$
\mathcal{V}_{\text{ac}}(t) = \frac{h}{ e \pi  w} \Big[\sum_{m=-\infty}^\infty \frac{1}{ 1 + (t- 2m \pi/\Omega)^2/w^2 } \Big] + C.
$
Here the offset $C$ is chosen to satisfy $\overline{\mathcal{V}_{\text{ac}}(t)}=0$.
The temperatures $\theta_L$ and $\theta_R$ are fixed while the average chemical potential $\mu\equiv (\mu_L+\mu_R)/2$ and the AC frequency $\Omega$ are tuned, choosing the DC voltage bias $(\mu_L-\mu_R)/e$ that maximizes the generated power (see SI for details). 
Fig.~\ref{fig:eng-chiral}(a) shows that when the AC driving becomes nonadiabatic,  $\hbar\Omega > \Gamma$, the efficiency becomes significantly enhanced with respect to the static cases $\hbar\Omega=0$.
Notably, when the average chemical potential aligns with the resonance level, the driving can even make the system to operate as a heat engine while it does not in the static case, hence realizing a photoassisted thermoelectric engine [see inset in Fig.~\ref{fig:eng-chiral}(a)].
This is due to the electron-hole asymmetry dynamically induced by the Levitons.
For an AC voltage of a sine-wave form, which does not break electron-hole symmetry, such effect is not observed.
As the frequency increases, the total transmission probability $\mathcal{T}$, see Fig.~\ref{fig:eng-chiral}(b), shows subpeaks determined by photon-assisted transmissions.
For Levitons, the photoassisted transition probability  $|a_n|^2$, shown in the inset of Fig.~\ref{fig:eng-chiral}(b), becomes asymmetric for photon absorption and emission. The analytic expression of $a_n$ is written in the SI, which is experimentally verified by quantum tomography~\cite{jullien2014quantum}.

Remarkably, the efficiencies in the AC driven cases become larger than the Carnot bound. This behavior is accompanied by the negative entropy production $\overline{\dot{S}^{(C)}} $ given by the Clausius relation as shown in Fig.~\ref{fig:eng-chiral}(c). We emphasize that the efficiency enhancement over the Carnot efficiency is not restricted to the AC voltage of Lorentzian pulses; e.g. we also observe efficiencies beyond Carnot bound for sinusoidal signals.

\begin{figure}[t]
  \centering
  \includegraphics[width=0.95\columnwidth]{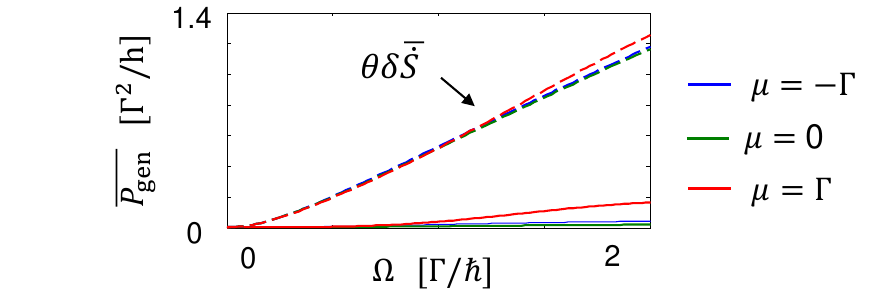}
  \caption{\textbf{Power generation from isothermal baths, $\theta_L=\theta_R=\theta$.}
    The power generation is allowed upto $\theta \delta \overline{\dot{S}}$ (dashed lines) when the AC driving induces additional entropy production, $\delta \overline{\dot{S}}>0 $.    Here $k_B \theta = 0.25 \Gamma$. 
}
  \label{fig:Pgen}
\end{figure}

This seeming violation of the second law of thermodynamics is understood from the fact that the Shannon entropy production Eq.~(\ref{eq:dS-f}) is always nonnegative.
Figure~\ref{fig:ddS}(a) shows that the deviation of the entropy production from the Clausius equality, $\delta \overline{\dot{S}}$, grows as the frequency increases.
Up to $\Omega \sim k_B \theta_L/\hbar$, the deviation $\delta \overline{\dot{S}}$ increases quadratically with respect to $\Omega$, when $w \Omega$ is fixed, independently of the values of $\mu$ as predicted by Eq.~(\ref{eq:ddS}) because the photon number uncertainty is determined only by the AC voltage driving protocol (see the inset).
The positivity of the Shannon entropy production provides a new upper bound $\eta_C^{\text{(gen)}}= \eta_C + \theta_R \delta \overline{\dot{S}}/|\overline{I_h^L}|$ of the efficiency, rather than the Carnot value. Importantly, $\eta_C^{(\text{gen})}$ is not universal unlike $\eta_C$ and depends on the details of the nonequilibrium AC potential. Therefore, there is considerable room to tailor arbitrarily large efficiencies (limited by energy conservation) in AC driven quantum chiral conductors.
The upper bound increases as the frequency increases because $\delta \overline{\dot{S}}$, hence the nonequilibrium effect, is enhanced [see Fig.~\ref{fig:ddS} (b)].
The anomalous efficiency enhancement by the AC driving becomes stronger when the Lorentzian pulse is more squeezed in the period, as shown by  Fig.~\ref{fig:ddS}(c). This is as expected, because when the pulse is more squeezed,
the energy uncertainty $\delta n\, \hbar\Omega=e(\overline{\mathcal{V}_{\text{ac}}^2})^{1/2}$ becomes larger, hence $\delta \overline{\dot{S}}$ is enhanced.

Fig.~\ref{fig:Pgen} shows that our engine extracts work even when the temperatures of both reservoirs are equal, $\theta_L=\theta_R=\theta$, violating the traditional Kelvin-Planck statement~\cite{planck2013treatise} of the second law of thermodynamics.
  The power is generated when an electric current is produced by the AC driving and when a electric voltage of a value  smaller than a stopping voltage (i.e. a voltage which stops the current) is applied against the current.
  We find that the electric pumping current has a simple expression
  when $\delta n \, \hbar\Omega \ll k_B \theta \ll \Gamma$ (see SI for the derivation)
\begin{equation}
  \label{eq:IePump}
  \left.\overline{I_e^R}\rule{0cm}{0.2cm}\right|_{V_L=V_R}
  = \frac{e (\delta n \, \hbar \Omega)^2 }{2h } 
  \frac{d |t_{\text{st}}|^2}{d \mathcal{E}}\Big|_{\mathcal{E}=\mu} .
\end{equation}
This is a universal relation which suggests that the power generation beyond the Carnot limit is possible for any local scatterer of energy-dependent transmission and any AC voltage with nonvanishing fluctuation $\overline{\mathcal{V}_{\text{ac}}^2}$. Accordingly, we observe the efficiency enhancement beyond the Carnot limit for the AC voltage of sine-wave signal (not shown) or for a scatterer realized by a quantum point contact~\cite{buttiker1990quantized} (see SI).

One may define an effective temperature~\cite{alicki2015non,agarwalla2017quantum,latune2021roles} $\theta_{\text{eff}}$ of the left reservoirs, by comparing Eq.~(\ref{eq:IePump}) with a thermoelectric current in the same situation but substituting the left reservoirs by a static reservoir of temperature $\theta_{\text{eff}}$.
In this situation, the thermoelectric current satisfies 
$\overline{I_e^{R}}|_{V_L=V_R}=[e \pi^2 k_B^2 \theta (\theta_{\text{eff}}-\theta)/3h]  (d |t_{\text{st}}|^2/d \mathcal{E})|_{\mathcal{E}=\mu}$~\cite{benenti2017fundamental}, 
in the same condition used to derive Eq.~(\ref{eq:IePump}), namely $|\theta_{\text{eff}}-\theta|  \ll \theta \ll \Gamma/k_B$.
This suggests the effective temperature being $ \theta_{\text{eff}} = \theta+ 3(\delta n \, \hbar \Omega)^2/(2 \pi^2 k_B^2 \theta)$, which is manifestly positive.
  The effective temperature, which measures the broadening of a Fermi-Dirac-like distribution~\cite{pekola2004limitations}, is enhanced due to the AC voltage which rearranges the distribution of
electrons in energy in a more uncertain way due to finite  $\delta n$, without injecting any net work over a period on the system. However, caution is needed; such effective temperature does not  describe the photoassisted distribution $f_{\alpha}^{(\text{out})}(\mathcal{E})$ well, because it strongly departs from a Fermi-Dirac distribution when the AC driving becomes nonadiabatic, see Supplementary Figure 3 in SI.
Furthermore, the photoassisted electrons have coherence, i.e., the off-diagonal element in the density matrix in the energy basis, which is known to induce nontrivial effect in the current noise~\cite{battista2014energy} and cannot be described by the effective temperature.
This coherence does not play any role on the time-averaged values of charge and heat current considered here.
But it may significantly affect other functionalities of the heat engine, such as time-resolved currents and the thermodynamic uncertainty relation~\cite{kheradsoud2019power}.

Alternatively, the role of the AC voltage can be interpreted as a nonequilibrium demon~\cite{sanchez2019nonequilibrium}.
Here, the  nonequilibrium demon (AC driving) induces additional entropy production by rearranging the distribution of electrons in energy in a more uncertain way, while satisfying the demon condition: no injection of energy $\overline{P_{\text{in}}}=0$.
Thus, the entropy production deviates from the Clausius equality, and the power generation is allowed up to the upper bound $\theta \delta {\dot{S}}$.
In contrast to the setups of Ref.~\cite{sanchez2019nonequilibrium},
our setup does not need a fine tuning for the demon condition $\overline{P_{\text{in}}}=0$; the condition is satisfied regardless of the AC voltage profile.

\vspace{2mm}
\textbf{Discussion}

This work demonstrates that an AC driven chiral conductor can exhibit efficiencies beyond the Carnot limit due to the negative entropy production when the Clausius relation is assumed. To amend the apparent violation of the second law of thermodynamics we employ a positively defined entropy production based on the Shannon entropy applied for the incoming and outgoing electronic distribution functions of the reservoirs. Interestingly, we find that the deviation of entropy production from the Clausius relation is given by the photon number uncertainty of the AC driving.
Chiral transport is crucial for the efficiency enhancement.
Nonchiral conductors do not exhibit efficiencies beyond the Carnot limit due to the finite injection energy which diminishes the generated power.
The regime for achieving efficiency beyond the Carnot limit, $\Omega> \Gamma/\hbar$, is realistic because with  the state-of-the-art fast AC voltage of Lorentzian pulses of width $w=$ 15 ps  with AC period $2\pi/\Omega= $ 166 ps (or sinusoidal signal with AC period of 42 ps)~\cite{Leviton2013}, the regime $\Omega> \Gamma/\hbar$ is approachable as long as the level broadening is small enough as $\Gamma<$ 0.025 meV (or $\Gamma<$ 0.1 meV for the sinusoidal signal), which can be tuned with the tunnel coupling via finger gates.

\textbf{Data availability.} The authors declare that all data supporting the findings of this study are available within the article. 

\textbf{Reference.}

%

\vspace{2mm}

\textbf{Acknowledgements}

We acknowledge support from Grants No. MAT2017-82639 and No. 
PID2020-117347GB-I00 funded by MCIN/AEI/10.13039/501100011033, No. MDM2017-0711 funded by MINECO/AEI/FEDER María de Maeztu Program for Units of Excellence, and No. PDR2020-12 funded by GOIB.
SR also acknowledges partial support from National Research Foundation of Korea (Grant No. 2021R1A6A3A03040076).

\vspace{2mm}
\textbf{Author Contributions Statement}
  All authors, S. R., R. L., L. S., and D. S. conceived the project.
  S. R. performed the calculations.
  All authors discussed the results and made significant contributions to the manuscript.

\vspace{2mm}
\textbf{Competing interests:} The authors declare no competing interests.

\clearpage
\newpage
\setcounter{page}{1}
\setcounter{equation}{0}
\setcounter{figure}{0}
\renewcommand{\figurename}{Supplementary Figure}
\renewcommand{\tablename}{Table}
\renewcommand{\refname}{Supporting Reference}
\renewcommand{\thetable}{S\arabic{table}}
\renewcommand{\thefigure}{\arabic{figure}}
\def\@cite#1{\textsuperscript{S#1}}
\renewcommand{\thesubsection}{\Roman{subsection}}
\renewcommand{\theequation}{S\arabic{equation}}
\onecolumngrid
\begin{center}{\textbf{\large Supplementary Information: Beating Carnot efficiency with periodically driven chiral conductors}\\
    \vspace{2mm} 
    Sungguen Ryu, Rosa L\'{o}pez, Lloren\c{c} Serra, and David S\'{a}nchez}\\
  {\it Instituto de F\'{i}sica Interdisciplinar y Sistemas Complejos IFISC (CSIC-UIB), E-07122 Palma, Spain}\end{center}
\vspace{1cm}
\twocolumngrid

\section{Phototransition amplitude $a_n$ }

\begin{figure}[h]
  \centering
  \includegraphics[width=\columnwidth]{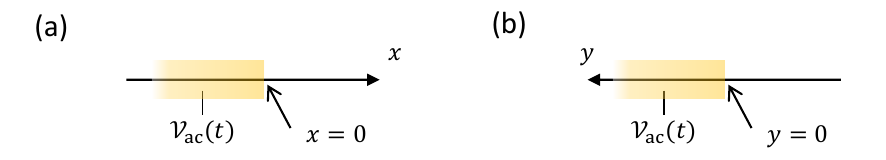}
  \caption{Chiral channel with AC driven contact. Panel (a) is when an electron propagates out of the AC driven contact and (b) when an electron propagates into the contact.
    }
  \label{fig:ac-bd}
\end{figure}
For completeness, we summarize the derivation~\cite{keeling2006minimal,dubois2013integer} that an electron propagating in a chiral channel out of an AC driven contact [see Supplementary Figure~\ref{fig:ac-bd} (a)] absorbs (or emits) $|n|$ photons with the transition amplitudes $a_n$ of Eq.~(\ref{eq:an}), for $n>0$ ($n<0$).

When an electron of initial energy $\mathcal{E}$ propagates from the AC driven contact [see Supplementary Figure~\ref{fig:ac-bd} (a)], its wave function solution of the time-dependent Schr\"{o}dinger equation is 
\begin{equation}
  \psi_{\mathcal{E}} (x,t) = e^{-i \mathcal{E} (t-x/v)}  
   \Big[\Theta(-x)e^{i \phi_{\text{ac}}(t)}
    +\Theta(x) e^{i \phi_{\text{ac}}(t-x/v)} \Big].
\end{equation}
Here, $v$ is propagation velocity for the chiral channel,
$\phi_{\text{ac}}(t) = - e\int_{-\infty}^t dt' \mathcal{V}_{\text{ac}}(t')/\hbar$, and $\Theta(x)$ is 0 for $x<0$ and 1 otherwise.
This solution can be verified by plugging it into the Schr\"{o}dinger equation for the Hamiltonian of linear dispersion, $ [i\hbar(\partial/\partial t + v\partial/\partial x) -e \mathcal{V}_{\text{ac}}(t)\Theta(-x)]\psi_{\mathcal{E}}(x,t)=0$. When an electron is in the contact, $x<0$, the phase factor $e^{i \phi_{\text{ac}}(t)}$ does not induce any change of energy as it is a spatially independent phase shift.
When an electron is out of the contact, $x>0$, the spatially dependent phase factor $e^{i\phi_{\text{ac}}(t-x/v)}$ describes the electronic energy change due to the photoassisted transition.
By definition of $a_n$ in Eq.~(\ref{eq:an}), $e^{i\phi_{\text{ac}}(t-x/v)} = \sum_n a_n e^{-i n \Omega (t-x/v)}$, thus the transition amplitude with which the electron changes its energy by $n\hbar\Omega$ is $a_n$.

We also remark the case when an electron propagates into the AC driven contact [see Supplementary Figure~\ref{fig:ac-bd} (b)], which is used for the results of nonchiral conductors shown below.
The wave function solution of the time-dependent Schr\"{o}dinger equation with initial energy $\mathcal{E}$ is 
\begin{equation}
  \psi_{\mathcal{E}} (y,t) = e^{-i \mathcal{E} (t-y/v)}  
   \Big[\Theta(-y)
    +\Theta(y) e^{i \{\phi_{\text{ac}}(t)- \phi_{\text{ac}}(t-y/v)\}} \Big].
\end{equation}
This can be verified by plugging it into the Schr\"{o}dinger equation
$ [i\hbar(\partial/\partial t + v\partial/\partial y) -e \mathcal{V}_{\text{ac}}(t)\Theta(y)]\psi_{\mathcal{E}}(y,t)=0$.
According to the solution and $e^{-i\phi_{\text{ac}}(t-y/v)} = \sum_n a_n^* e^{i n \Omega (t-y/v)}=\sum_n a_{-n}^* e^{-i n \Omega (t-y/v)}$, the transition amplitude with which the electron changes its energy by $n\hbar\Omega$ is $a_{-n}^*$.

\section{Unitarity of Floquet Scattering matrix of Eq.~(\ref{eq:SF})}

Here we show that the Floquet scattering matrix of Eq.~(\ref{eq:SF}) satisfies the unitarity conditions, $\sum_{n \alpha} |\mathcal{S}_{\alpha \beta}(\mathcal{E}_n , \mathcal{E})|^2 =1$ and $\sum_{n \beta} |\mathcal{S}_{\alpha \beta}(\mathcal{E} , \mathcal{E}_{-n})|^2 =1$.

The first condition is satisfied as,
\begin{align}
  \sum_{n \alpha} |\mathcal{S}_{\alpha L}(\mathcal{E}_n, \mathcal{E})|^2
  &= \sum_n |a_n|^2 \Big(|t_{\text{st}}(\mathcal{E}_n)|^2 + |r_{\text{st}}(\mathcal{E}_n)|^2 \Big)  \nonumber \\
  &= \sum_n |a_n|^2 =1, \\
\sum_{n \alpha} |\mathcal{S}_{\alpha R}(\mathcal{E}_n, \mathcal{E})|^2
  &=  |t'_{\text{st}}(\mathcal{E})|^2 + |r'_{\text{st}}(\mathcal{E})|^2= 1.
\end{align}
     
The second condition is satisfied as,
\begin{align}
  \sum_{n \beta} |\mathcal{S}_{L \beta}(\mathcal{E}, \mathcal{E}_{-n})|^2
  &= \sum_n |a_n|^2 |r_{\text{st}}(\mathcal{E})|^2 + |t'_{\text{st}}(\mathcal{E})|^2  \nonumber\\
  &= |r_{\text{st}}(\mathcal{E})|^2 + |t'_{\text{st}}(\mathcal{E})|^2 =1, \\
  \sum_{n \beta} |\mathcal{S}_{R \beta}(\mathcal{E}, \mathcal{E}_{-n})|^2
  &=  \sum_n |a_n|^2 |t_{\text{st}}(\mathcal{E})|^2 + |r'_{\text{st}}(\mathcal{E})|^2  \nonumber \\
  &= |t_{\text{st}}(\mathcal{E})|^2 + |r'_{\text{st}}(\mathcal{E})|^2
   = 1. 
\end{align}

\section{Derivation of $\braket{n}=0$}

Here we prove that the mean number of photons $\braket{n}=\sum_n n |a_n|^2$ involved in the photoassisted transitions by the AC driving is zero.

Let $a(t)$ be the Fourier transform of $a_n$,
\begin{align}
  a(t) &\equiv \sum_n a_n e^{-i n\Omega t},   \label{eq:a-t} \\
  &= \exp\left[-i \frac{e}{\hbar} \int_{-\infty}^t dt'\, \mathcal{V}_{\text{ac}}(t')\right]. \label{eq:a-t2}
\end{align}
The mean number $\braket{n}$ is expressed in terms of $a(t)$ as
\begin{equation}
  \sum_n n |a_n|^2
  = \frac{i}{2\pi}   \int_0^{2\pi/\Omega} dt\,  \frac{\partial a}{\partial t} a^*(t)\;.
\end{equation}
This can be verified when using Eq.~(\ref{eq:a-t}) and $\int_0^{2\pi/\Omega} dt \, e^{i (n-n')\Omega t} =(2\pi/\Omega) \delta_{nn'} $ for integers $n$ and $n'$.
Using that $|a(t')|^2=1$, 
\begin{equation}
  \frac{\partial a}{\partial t} a^*(t)
  = \frac{1}{a(t)}\frac{\partial a}{\partial t} 
  = \frac{\partial \ln a}{\partial t}
  = -i \frac{e \mathcal{V}_{\text{ac}}(t)}{\hbar}.
  \label{eq:d-ln-at}
\end{equation}
In the last equality we used Eq.~(\ref{eq:a-t2}). Hence, the mean number $\braket{n}$ is determined by the time-averaged AC voltage $\overline{\mathcal{V}_{\text{ac}}}$
\begin{equation}
  \braket{n}
  = \frac{e \overline{\mathcal{V}_{\text{ac}}(t)}}{\hbar\Omega }.
\end{equation}
This vanishes because $\overline{\mathcal{V}_{\text{ac}}}=0$.

\section{Photon number uncertainty $\delta n$ in terms of AC voltage profile $\mathcal{V}_{\text{ac}}(t)$}

Here,  we derive that the photon number uncertainty $\delta n$ is equal to the ratio between the root mean square of AC voltage and the photon energy quantum
\begin{equation}
  \delta n = \frac{e \big\{\overline{\mathcal{V}_{\text{ac}}^2(t)}\big\}^{1/2}}{\hbar\Omega}.
  \label{eq:dn-Vac}
\end{equation}

\begin{proof}
  We start by the fact that  $\braket{n^2}$ is written in terms of $a(t)$ (see Eq.~(\ref{eq:a-t})-~(\ref{eq:a-t2})) as
  \begin{equation}
    \sum_n n^2 |a_n|^2 = - \frac{1}{2\pi \Omega} \int_0^{2\pi/\Omega} dt\, \frac{\partial^2 a}{\partial t^2} a^*(t).
  \end{equation}
This can be verified when using Eq.~(\ref{eq:a-t}) and $\int_0^{2\pi/\Omega} dt \, e^{i (n-n')\Omega t} =(2\pi/\Omega) \delta_{nn'} $ for integers $n$ and $n'$.
Then, we use the integration by parts,
\begin{align}
  \sum_n n^2 |a_n|^2
  &= -\frac{1}{2\pi \Omega} \Big[ \frac{\partial a}{\partial t}a^*(t) \Big]_0^{2\pi/\Omega} \nonumber \\
  &\qquad + \frac{1}{2\pi \Omega}   \int_0^{2\pi/\Omega} \hspace{-0.2cm}dt\, \frac{\partial a}{\partial t} \frac{\partial a^*}{\partial t}\;.
\end{align}
The first term vanishes using Eq.~(\ref{eq:d-ln-at}). We relate the second term to the AC voltage using $|a(t)|^2=1$ and 
\begin{equation}
  \label{eq:1}
  \frac{\partial a^*}{\partial t}
  = \frac{\partial }{\partial t} \frac{1}{a}
  = - \frac{1}{a^2} \frac{\partial a}{\partial t}.
\end{equation}
Using this and (\ref{eq:d-ln-at}), we obtain
\begin{equation}
  \label{eq:3}
  \sum_n n^2 |a_n|^2
  = \frac{1}{2\pi \Omega} \int_0^{2\pi/\Omega} dt\,
  \Big(\frac{e \mathcal{V}_{\text{ac}}(t)}{\hbar}  \Big)^2.
\end{equation}
Then Eq. (\ref{eq:3})  is equal to Eq.~(\ref{eq:dn-Vac}) squared.
\end{proof}

\section{Derivation of Eq.~(\ref{eq:ddS})}

Here we derive the deviation of entropy from the Clausius relation, Eq.~(\ref{eq:ddS}), in the regime of small biases $k_B|\theta_L -\theta_R|, \, k_B|\mu_L-\mu_R| \ll k_B \theta_L$,
and small energy uncertainty induced by the AC voltage, $\delta n \,\hbar\Omega \ll k_B\theta_L$.

First, we approximate the outgoing distribution $f^{\text{(out)}}_{\alpha}(\mathcal{E})$ in the small driving frequency. We use the relation of the outgoing distribution to the ingoing distribution in terms of the Floquet scattering matrix,
\begin{align}
  f^{\text{(out)}}_\alpha (\mathcal{E})
  &= \sum_{\beta n} |\mathcal{S}_{\alpha \beta}(\mathcal{E}, \mathcal{E}_{-n})|^2  f_{\beta}(\mathcal{E}_{-n}).
    \label{eq:fout-in}
\end{align}
Due to the small energy uncertainty $\delta n \,\hbar\Omega \ll k_B\theta_L$, the ingoing distribution shifted by energy quanta (much smaller than thermal broadening) is similar to the original distribution. Therefore, we approximate
\begin{equation}
  \label{eq:fin-exp}
  f_{\beta}(\mathcal{E}_{-n})
  = f_{\beta}(\mathcal{E}) - n \, \hbar\Omega f'_{\beta}(\mathcal{E})
  + \frac{(n\, \hbar\Omega)^2}{2} f''_{\beta}(\mathcal{E}).
\end{equation}
Here $f_\beta'(\mathcal{E})$ and $f_\beta''(\mathcal{E})$ are the first and second derivative of the ingoing distribution $f_\beta(\mathcal{E})$.
Then, plugging Eq.~(\ref{eq:fin-exp}) into Eq.~(\ref{eq:fout-in}), we obtain an approximation for the output distribution,
\begin{align}
  \label{eq:2}
  f^{\text{(out)}}_{\alpha} (\mathcal{E})
  &=\sum_{\beta n} |\mathcal{S}_{\alpha \beta}(\mathcal{E}, \mathcal{E}_{-n})|^2 f_\beta (\mathcal{E}) \nonumber \\
  &\quad -\sum_{\beta n} n |\mathcal{S}_{\alpha \beta}(\mathcal{E}, \mathcal{E}_{-n})|^2 \, \hbar \Omega \, f'_\beta(\mathcal{E})    \nonumber \\
  &\quad +\frac{1}{2} \sum_{\beta n} n^2 |\mathcal{S}_{\alpha \beta}(\mathcal{E}, \mathcal{E}_{-n})|^2 \,(\hbar \Omega)^2 \, f''_\beta(\mathcal{E})\;.
\end{align}
We use Eq.~(\ref{eq:SF}) for the Floquet scattering matrix
and sum over the photon number. Using $\braket{n}=0$, we have
\begin{align}
  f^{\text{(out)}}_{\alpha} (\mathcal{E})
  &=f^{\text{(out,st)}}_\alpha (\mathcal{E}) \nonumber \\
  &\quad +\frac{1}{2}  (\delta n)^2 \, |\mathcal{S}_{\alpha L}^{\text{(st)}}(\mathcal{E})|^2\, f''_L(\mathcal{E})\,  (\hbar \Omega)^2\;.
    \label{eq:f-out-omg}
\end{align}
Here, $f^{\text{(out,st)}}_\alpha(\mathcal{E})\equiv \sum_{\beta} |\mathcal{S}_{\alpha \beta}^{\text{(st)}}(\mathcal{E})|^2 f_{\beta}(\mathcal{E})$ is the outgoing distribution in the static case. 

From Eq.~(\ref{eq:f-out-omg}), we use that the second term is much smaller than the first term, due to the condition $\delta n \, \hbar \Omega \ll k_B \theta_L$, and we approximate the Shannon entropy of the outgoing distribution up to leading order in $\Omega$, 
\begin{align}
  &\sigma[ f^{\text{(out)}}_\alpha (\mathcal{E})]
  =   \sigma[ f^{\text{(out,st)}}_\alpha (\mathcal{E})] \nonumber \\
  &\qquad +\sigma'[f^{\text{(out,st)}}_\alpha (\mathcal{E})] \, \frac{(\delta n)^2}{2} \, |\mathcal{S}_{\alpha L}^{\text{(st)}}
  (\mathcal{E})|^2 \, f''_L(\mathcal{E}) \,  (\hbar\Omega)^2\;.
\end{align}
Here $\sigma'[x]\equiv d \sigma[x]/dx$.
Then, we obtain the approximation of the entropy production with the condition $\delta n \, \hbar \Omega \ll k_B \theta_L$,
\begin{align}
  \label{eq:dS-smalldE}
  &\overline{\dot{S}}
  =   \overline{\dot{S}^{\text{(st)}}} \nonumber \\
  &+ \sum_\alpha \frac{(\delta n)^2}{2h} \int d \mathcal{E}\,
  k_B\, \sigma'[f^{\text{(out,st)}}_\alpha(\mathcal{E})]\,
  |\mathcal{S}_{\alpha L}^{\text{(st)}}(\mathcal{E})|^2\, f''_L(\mathcal{E}) 
  (\hbar\Omega)^2,
\end{align}
where $\overline{\dot{S}^{\text{(st)}}}$ is the time-averaged entropy production in the static case.

Now, we apply the condition of the small biases to Eq.~(\ref{eq:dS-smalldE}).
In the leading order of the small biases,
$  \overline{\dot{S}^{\text{(st)}}}= \sum_\alpha \overline{I_h^\alpha}/\theta_\alpha$ (see below) and $\sigma'[f^{\text{(out,st)}}_\alpha(\mathcal{E})] \approx \sigma'[f^{\text{(in)}}_L(\mathcal{E})]$. Using  $\sigma'[f^{\text{(in)}}_L(\mathcal{E})] = (\mathcal{E}-\mu_L)/(k_B \theta_L)$ and $\sum_\alpha |\mathcal{S}^{(\text{st})}_{\alpha L}(\mathcal{E})|^2 = 1$, we obtain the deviation of the entropy production from the Clausius relation, in the case of small biases and small energy uncertainty
\begin{equation}
  \delta \overline{\dot{S}}
  =   \frac{(\delta n \, \hbar \Omega)^2}{2h \theta_L} \int d \mathcal{E}\, f''_L(\mathcal{E}) (\mathcal{E}-\mu_L).
\end{equation}
This yields Eq.~(\ref{eq:ddS}), because using the integration by parts, $\int d \mathcal{E}\, f''_L(\mathcal{E}) (\mathcal{E}-\mu_L) = -\int d \mathcal{E} \, f'_L(\mathcal{E})=1$.

For completeness, we show the derivation for $  \overline{\dot{S}^{\text{(st)}}} = \sum_\alpha \overline{I_h^\alpha}/\theta_\alpha$ in the leading order of the small biases. The outgoing distribution in the static case is expanded for small biases as,
\begin{align}
  f^{\text{(out,st)}}_\alpha (\mathcal{E})
  &=  \sum_\beta |S^{\text{(st)}}_{\alpha \beta}(\mathcal{E})|^2
    f_{\beta}(\mathcal{E}) \nonumber \\
  &= f_\alpha (\mathcal{E}) + \sum_\beta  |S^{\text{(st)}}_{\alpha \beta}(\mathcal{E})|^2 \{f_\beta (\mathcal{E}) -f_\alpha (\mathcal{E})\}\;.
\end{align}
Then, Shannon entropy of the outgoing distribution is expanded,
\begin{align}
  &\sigma[ f^{\text{(out,st)}}_\alpha (\mathcal{E})]
  = \sigma[ f_\alpha (\mathcal{E})] \nonumber\\
 &\qquad + \sigma'[ f_\alpha (\mathcal{E})] \sum_\beta  |S^{\text{(st)}}_{\alpha \beta}(\mathcal{E})|^2 \{f_\beta (\mathcal{E}) -f_\alpha (\mathcal{E})\}\;.
\end{align}
Using $\sigma'[f_\alpha (\mathcal{E})]= (\mathcal{E}-\mu_\alpha)/\theta_\alpha$, the entropy production is found to be,
\begin{align}
  \overline{\dot{S}^{\text{(st)}}}
  = \sum_{\alpha \beta }\frac{1}{h} \int d \mathcal{E}\,
  \frac{\mathcal{E}-\mu_\alpha}{\theta_\alpha}
  |S_{\alpha \beta}^{\text{(st)}}(\mathcal{E})|^2
  \{f_\beta (\mathcal{E}) - f_\alpha (\mathcal{E})\}\;.
\end{align}
This equals $\sum_\alpha \overline{I_h^\alpha}/\theta_\alpha$.

\section{Choice of DC voltage bias to maximize the power }

Here we discuss the choice of DC voltage bias which maximizes the generated power, for given temperatures of the reservoirs and average chemical potential $\mu$. Then, this bias is used for obtaining the results of Figs.~\ref{fig:eng-chiral}, \ref{fig:ddS} and \ref{fig:Pgen}.
We followed the approach of the linear thermoelectricity~\cite{benenti2017fundamental}, deriving response coefficients in the linear regime, namely when the thermal and voltage biases are small compared to the average temperature, $k_B |\Delta \theta|, |e\Delta V| \ll k_B\theta$, where $\Delta \theta \equiv \theta_L-\theta_R$, $\Delta V \equiv (\mu_L-\mu_R)/e$, $\theta \equiv (\theta_L+\theta_R)/2$.

To expand the generated power $-\Delta V \overline{I_e^R}$, ($\overline{P_{\text{in}}}=0$ for the setup of chiral conductors) we consider the linear regime for the current up to the first order of the biases $\Delta V$ and $\Delta \theta$. Using Eq.~(\ref{eq:I-e}) and expanding the Fermi distributions in the small biases, we obtain 
$\overline{I_e^R}
  = \overline{I_e^R}|_{\Delta V =\Delta \theta =0} 
    + G \Delta V +L \Delta \theta$,
 \begin{eqnarray}
  G &=& \frac{e}{h} \int d \mathcal{E} \, \frac{\mathcal{T}(\mathcal{E}) + \mathcal{T}'(\mathcal{E}) }{2}
      \big(-f'(\mathcal{E}) \big),  \\
      L &=& \frac{1}{h} \int d \mathcal{E} \,  \frac{\mathcal{T}(\mathcal{E}) + \mathcal{T}'(\mathcal{E}) }{2}
      \big(-f'(\mathcal{E}) \big) \frac{\mathcal{E}-\mu}{\theta}. 
\end{eqnarray}    
Here, $G$ and $L$ are the response coefficients of the time-averaged electrical and thermoelectrical currents.
 $\overline{I_e^R}|_{\Delta V =\Delta \theta =0} $ is the electric pump current in the absence of the biases. $f(\mathcal{E})$ is the Fermi-Dirac distribution of the average temperature $\theta = (\theta_L+\theta_R)/2$ and average chemical potential $\mu = (\mu_L+\mu_R)/2$.  Now, the generated power is in a quadratic form of $\Delta V$, hence generated power is maximal for the voltage bias
\begin{equation}
  \Delta V =- \frac{ L }{ 2 G} \Delta \theta
  - \frac{\overline{I_e^R}|_{\Delta V =\Delta \theta =0}}{2G}.
  \label{eq:dV-max-chiral}
\end{equation}

\section{Photoasisted transition amplitudes for the Lorentzian voltage bias}

For completeness, we give the photoassisted transition amplitudes $a_n$ in the chiral conductor driven by the Lorentzian voltage pulses
described in the main text,
which is obtained in Ref.~\cite{dubois2013integer},
\begin{equation}
\begin{aligned}
  a_n &= - e^{- n \Omega w} ( 1 -e^{-2\Omega w})\,,
        \quad (n \ge 0) \\
  a_{-1} &=  e^{-\Omega w}\,,  \\           
  a_n &= 0\,.\quad (n\le -2)
\end{aligned}
\end{equation}

\section{Detailed results for Fig.~\ref{fig:eng-chiral}--\ref{fig:ddS}}

\begin{figure}[h]
  \centering
  \includegraphics[width=\columnwidth]{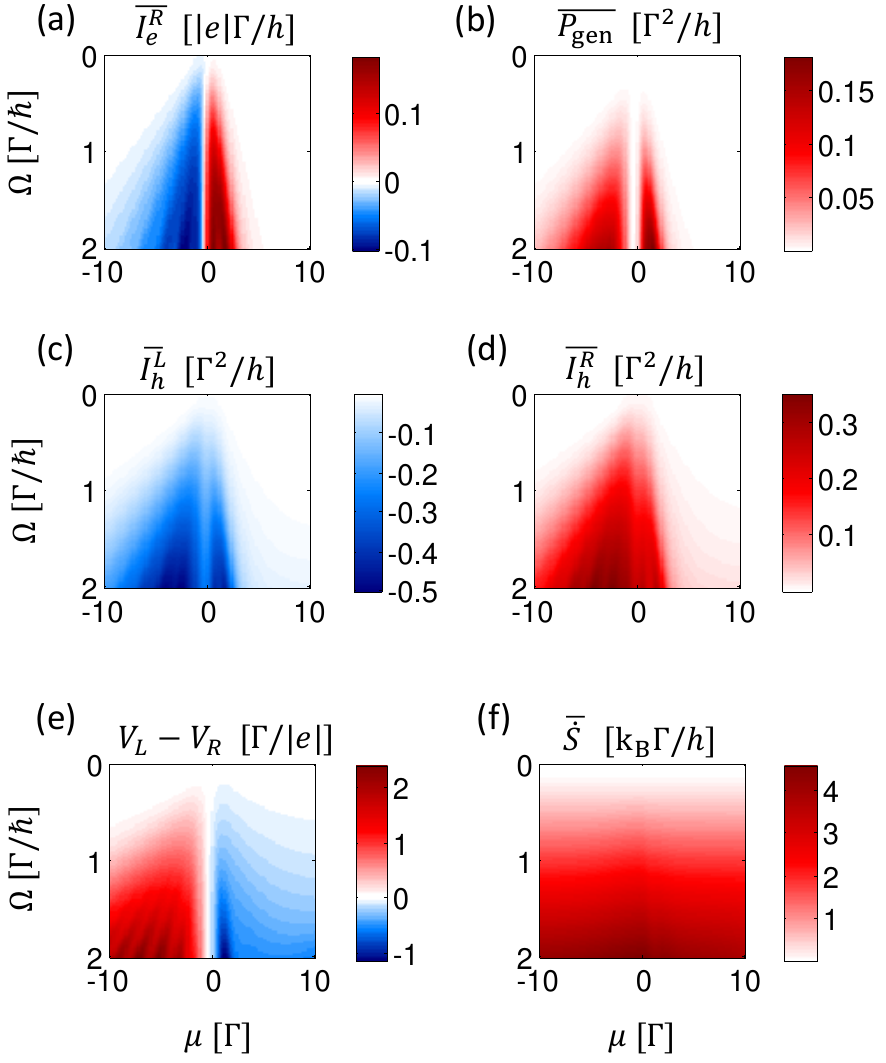}
  \caption{Detailed results used to obtain Figs.~\ref{fig:eng-chiral} and \ref{fig:ddS}, such as charge current (a), generated power (b), heat current into the left reservoir (c), heat current into the right reservoir (d),  DC bias voltage choices $(\mu_L-\mu_R)/e$ maximizing generated power [see Eq. (\ref{eq:dV-max-chiral})]  (e), and entropy production rate (f). In all the panels, horizontal axes are average chemical potential $\mu$ in units of $\Gamma$ and vertical axes are angular frequency of the ac voltage $\Omega$ in units of $\Gamma/\hbar$.}
  \label{fig:currents}
\end{figure}

Supplementary Figure~\ref{fig:currents} shows detailed results used to obtain Figs.~\ref{fig:eng-chiral}--\ref{fig:ddS}.

\begin{figure}[h]
  \centering
  \includegraphics[width=\columnwidth]{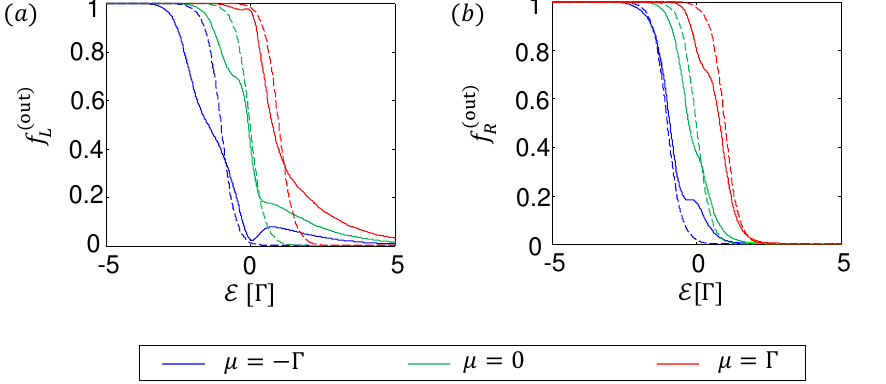}
  \caption{The outgoing distributions $f_L^{(\text{out})}$ (a) and $f_R^{(\text{out})}$ (b) for the situation of Fig.~\ref{fig:eng-chiral}.
      The results are shown for the AC driving frequencies of $\Omega=1$ (solid lines) and $\Omega=0$ (dashed lines) and the average chemical potentials $\mu=-\Gamma$ (blue), $\mu=0$ (green), and $\mu=\Gamma$ (red). }
  \label{fig:fout}
\end{figure}

Supplementary Figure~\ref{fig:fout} shows the outgoing distributions $f_{\alpha}^{(\text{out})}(\mathcal{E})$ in the situation of Fig.~\ref{fig:eng-chiral}, which depart from the Fermi-Dirac distribution for a fast AC driving.

\section{Engine with a quantum point contact}

\begin{figure*}[t]
  \centering
  \includegraphics[width=\textwidth]{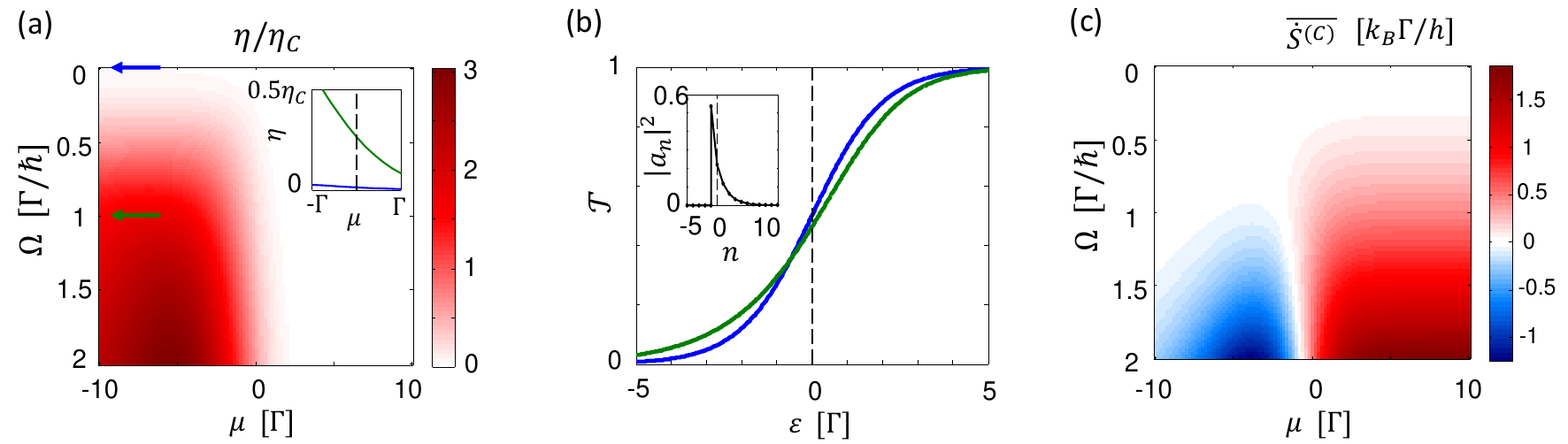}
  \caption{ Heat engine with a quantum point contact.  The same situation as in Fig.~\ref{fig:eng-chiral} substituting the quantum dot to a quantum point contact is considered.
    (a) Efficiency when tuning the driving frequency $\Omega$ and the average chemical potential $\mu\equiv (\mu_L+\mu_R)/2$ (measured from $E_0$). Inset: Plot near $\mu=0$ for $\Omega=0$ (blue) and $\Omega = \Gamma/\hbar$ (green) as indicated with arrows in the main panel.    
    (b) Total photoassisted transmission probability $\mathcal{T}$ for an electron of energy $\mathcal{E}$ (measured from $E_0$) incoming from the left reservoir for $\Omega=0$ (blue) and $\Omega = \Gamma/\hbar$ (green).
    Inset: the photoassisted transition probabilities $|a_n|^2$.
    (c) The entropy production rate assuming Clausius relation, $\overline{\dot{S}^{(C)}}$.
    Here, $(\theta_L+\theta_R)/2 = 0.25\Gamma/k_B$, $\theta_L-\theta_R = 0.1\theta$, $w= 0.05\times 2\pi/\Omega$, and the DC chemical potential bias is chosen for the maximal power generation.  }
  \label{fig:QPC-eta}
\end{figure*}

Here we present the results of our heat engine when implemented with a quantum point contact instead of the quantum dot. The transmission probability through the quantum point contact is modeled by a step-like transmission with a single channel,
$|t_{\text{st}}(\mathcal{E})|^2 = [1+\exp\{-(\mathcal{E}-E_0)/\Gamma \}]^{-1} $, where $E_0$ is the energy at the half transmission and $\Gamma$ is energy width of the step~\cite{buttiker1990quantized}, see Supplementary Figure~\ref{fig:QPC-eta}(b).
Supplementary Figure~\ref{fig:QPC-eta}(a) shows that the efficiency is also substantially enhanced, even beyond the Carnot limit.
Such anomalous efficiency enhacement is acompaned by the negative entropy production $\overline{\dot{S}^{(C)}}$ given by the Clausius relation as shown in Supplementary Figure~\ref{fig:QPC-eta}(c), similary to the case of the quantum dot.

We briefly discuss the differences between the case of QPC and the quantum dot. In QPC (quantum-dot) case, the efficiency is asymmetric (symmetric) for the average chemical potential $\mu$ around the half-transmission energy $E_0$ (resonance level). Namely, the efficiency is lower when $\mu>E_0$ when compared to the values for $\mu<E_0$.
This is because the electric conductance is larger for $\mu>E_0$, and hence i) the heat extracted from the hot reservoir is larger (due to increased thermal conductance) and ii) the choice of voltage bias for maximum power is lower (we recall that such voltage bias is determined by half of the thermoelectric coefficient divided by the electric conductance in the absence of the AC driving~\cite{benenti2017fundamental}).

\section{Derivation of Eq.~(\ref{eq:IePump})}

Here we derive Eq.~\eqref{eq:IePump}, the electric pumping current in the regime of $\hbar\Omega \ll k_B\theta\ll \Gamma$ and $\theta_L=\theta_R=\theta$. From Eq.~\eqref{eq:I-e}, the current becomes
\begin{align}
  \overline{I_e^R}_{\Delta V=\Delta \theta=0}
  &= e \int \frac{d \mathcal{E}}{h} \big\{ \mathcal{T}(\mathcal{E}) -\mathcal{T}' (\mathcal{E}) \big\} f(\mathcal{E}) \\
  &= e \int \frac{d \mathcal{E}}{h} \Big\{
    \sum_n |a_n|^2 \big|t_{\text{st}}(\mathcal{E}_n)\big|^2 -\big|t_{\text{st}}'(\mathcal{E})\big|^2  \Big\} f(\mathcal{E})
\end{align}
Using $|t_{\text{st}}'(\mathcal{E})|^2 = |t_{\text{st}}(\mathcal{E})|^2$, making an expansion in $\delta n \, \hbar\Omega/ (k_B \theta)$ up to second order, $\sum_n |a_n|^2=1$, $\sum_n n |a_n|^2 =0$, and integrating by parts,
we obtain 
\begin{equation}
  \overline{I_e^R}_{\Delta V=\Delta \theta=0}
  =\frac{e}{2} (\delta n \, \hbar \Omega)^2
  \int \frac{d \mathcal{E}}{h} \frac{ d |t_{\text{st}}|^2}{d \mathcal{E}} \big(-f'(\mathcal{E})\big) .  
\end{equation}
For temperatures much lower than the energy scale where transmission varies ($k_B\theta \ll \Gamma$), the Fermi function derivative can be replaced with a Delta funcion centered at $\mathcal{E}=\mu$ and then Eq.~(\ref{eq:IePump}) follows.

\section{Nonchiral conductors}

Here, we show that for nonchiral conductors, a large power injection $\overline{P}_{\text{in}}$ from the AC driving into the system diminishes the heat engine efficiency, in contrast to the case of chiral conductors. By analyzing the mean number of photons involved in the photoassisted scattering, we show that the AC power injection is positive and determined by Joule's law in the slow driving and zero temperature regime. We also show numerical results proving the efficiency decrease.

\begin{figure}[t]
  \centering
  \includegraphics[width=\columnwidth]{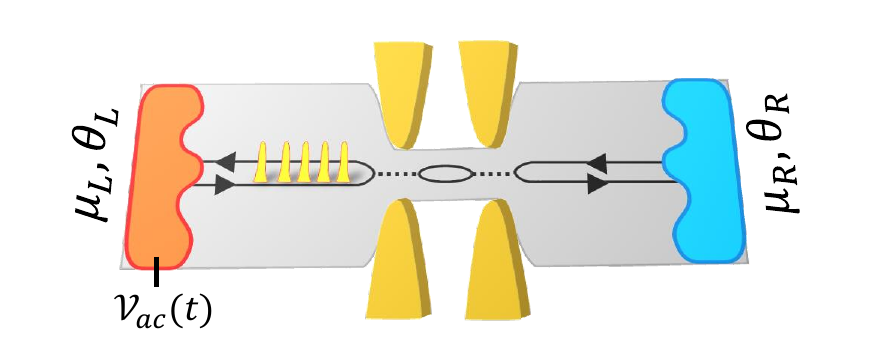}
  \caption{Setup of nonchiral conductor. In contrast to the chiral conductor, the left-going and right-going modes are not spatially separated.}
  \label{fig:setup-nonchi}
\end{figure}

Supplementary Figure~\ref{fig:setup-nonchi} shows the setup of nonchiral conductors driven by AC voltage bias.
In this case, the Floquet scattering matrix (including the photoassisted processes due to the AC voltage) becomes different from that of the chiral case, due to an additional process that an electron absorbs ($n>0$) or emits ($n<0$) $|n|$ photons when it is back scattered into the left contact with transition amplitude $a_{-n}^*$; see Supplementary Figure~\ref{fig:ac-bd} and related text. Therefore, the Floquet scattering matrix becomes
\begin{equation}
  \begin{aligned}
  S_{RL}(\mathcal{E}_n, \mathcal{E}) 
  &=  a_n t_{\text{st}}(\mathcal{E}_n), \\
  S_{LL} (\mathcal{E}_n, \mathcal{E})
  &= \sum_m a_m a^*_{-n+m} r_{\text{st}}(\mathcal{E}_m), \\
  S_{LR}(\mathcal{E}_n, \mathcal{E})
  &= a^*_{-n} t'_{\text{st}}(\mathcal{E}), \\
  S_{RR}(\mathcal{E}_n, \mathcal{E})
  &=  \delta_{n0} r'_{\text{st}}(\mathcal{E}).
  \end{aligned}
\label{eq:SF-nonchi-DriVol}                      
\end{equation}
Note that the reflection amplitude $S_{LL}(\mathcal{E}_n, \mathcal{E})$ from left input channel is determined by three steps rather than the two steps of the chiral case (see Eq.~(\ref{eq:SF})), due to the additional photon absorption/emission process which occurs when the electron finally enters the left reservoir.
As before, the Floquet scattering matrix satisfies the unitarity condition
$\sum_{n \alpha} |S_{\alpha \beta}(\mathcal{E}_n , \mathcal{E})|^2 =1$ and $\sum_{n \beta} |S_{\alpha \beta}(\mathcal{E} , \mathcal{E}_{-n})|^2 =1$.

The power injection from AC driving $\overline{P}_{\text{in}}$  does not vanish in the nonchiral setup, in contrast to the chiral setup. We show this by investigating the mean photon number involved in the photoassisted scattering. They fulfill
\begin{align}
  &\braket{n(\mathcal{E})}_{RL} +  \braket{n(\mathcal{E})}_{LL} 
  = \sum_n n |a_n|^2 \, |t_{\text{st}}(\mathcal{E}_n)|^2 \nonumber\\
  &\qquad +\sum_{n,m,l} n \{a_m a^*_{-n+m}r_{\text{st}}(\mathcal{E}_m)\}^* \, a_l a^*_{-n+l} \, r_{\text{st}}(\mathcal{E}_l)\;.  \label{eq:nL-nonchi}\\
  &\braket{n(\mathcal{E})}_{LR} +  \braket{n(\mathcal{E})}_{RR}
    =0\;.
\end{align}
The last term of Eq.~(\ref{eq:nL-nonchi}) is the result of interferences of all possible processes in which, e.g., an electron from the left reservoir first absorbs $m$ photons, reflects at the resonance level, and absorbs $m-n$ photons. The appearance of such term, contrarily to the case of chiral conductor, is key factor for finite AC power injection in the nonchiral setup.

To clarify the meaning of the last term of Eq.~(\ref{eq:nL-nonchi}), we change the variable $n$ to a new variable $ p\equiv -n+m$ in the last term in Eq.~(\ref{eq:nL-nonchi}),
\begin{align}
  &\braket{n(\mathcal{E})}_{RL} +  \braket{n(\mathcal{E})}_{LL}
  = \sum_n n |a_n|^2 \, |t_{\text{st}}(\mathcal{E}_n)|^2 \nonumber\\
  &\qquad -\sum_{p,m,l} p \, a_p a_{p+l-m}^* a_m^* a_l \, r_{\text{st}}^*(\mathcal{E}_m) \,  r_{\text{st}}(\mathcal{E}_l)  \nonumber \\
  &\qquad +\sum_{p,m,l} m \, a_p a_{p+l-m}^* a_m^* a_l\, r_{\text{st}}^*(\mathcal{E}_m)\, r_{\text{st}}(\mathcal{E}_l)\;.
    \label{eq:nL-nonchiral-1.5}
\end{align}
For the last term, we use
\begin{equation}
  \label{eq:sum-aa}
 \sum_{p}a_{p} a^*_{p+l-m} = \delta_{m,l}.
\end{equation}
This property follows from the fact that the left hand side is  the $(m-l)$-th Fourier coefficient of $e^{i \phi_{\text{ac}}(t)} e^{-i\phi_{\text{ac}}(t)}=1$.
Using Eq.~(\ref{eq:sum-aa}) the last term and the first term of the right hand side of Eq.~(\ref{eq:nL-nonchiral-1.5}) vanish due to $\braket{n}=0$.
Then, we obtain 
\begin{align}
  \braket{n(\mathcal{E})}_{RL} +  \braket{n(\mathcal{E})}_{LL}
  &= -\sum_{p,m,l} p\, a_{p} a^*_{p+l-m}  \nonumber \\ &\qquad  \times a^*_m a_l\, r^*_{\text{st}}(\mathcal{E}_m)\, r_{\text{st}}(\mathcal{E}_l)\;. \label{eq:nL-nonchi-2}                                                \end{align}
We utilize the fact that [$a(t)$ is the Fourier transform of the phototransition probabilities, see Eq.~(\ref{eq:a-t})] 
\begin{align}
  \sum_{p} p\, a_{p} a^*_{p+l-m}
  &= \frac{i}{2\pi}   \int_0^{2\pi/\Omega} dt\,  \frac{\partial a}{\partial t} a^*(t) e^{i(m-l) \Omega t} \nonumber \\
  &= \frac{1}{2\pi}   \int_0^{2\pi/\Omega} dt\,  \frac{e \mathcal{V}_{\text{ac}}(t)}{\hbar} e^{i(m-l) \Omega t}\;. \label{eq:sum-aa2}
\end{align}
The first equality can be verified when employing the definition of $a(t)$, Eq.~(\ref{eq:a-t}).
In the second equality, Eq.~(\ref{eq:d-ln-at}) is used.
Using Eqs.~(\ref{eq:nL-nonchi-2}), (\ref{eq:sum-aa2}), we obtain the mean number
\begin{equation}
  \braket{n(\mathcal{E})}_{RL} +  \braket{n(\mathcal{E})}_{LL}
  = -\frac{1}{2\pi}\int_0^{2\pi/\Omega} dt\,
  \frac{e \mathcal{V}_{\text{ac}}(t)}{\hbar} |\psi_r(t; \mathcal{E})|^2,
  \label{eq:nL-nonchi-3}
\end{equation}
where $\psi_r(t; \mathcal{E})\equiv \sum_l a_l r_{\text{st}}(\mathcal{E}_l) e^{-i l\Omega t}$ is the wave function immediately after the reflection at time $t$ for an electron of initial  energy $\mathcal{E}$.
The mean number of Eq.~(\ref{eq:nL-nonchi-3}) is generally nonzero and contributes to the AC power injection as explicitly demonstrated for the slow driving and zero temperature limits.

In the slow driving regime, $\hbar \Omega \ll \Gamma$, 
we further simplify Eq.~(\ref{eq:nL-nonchi-3}).
Expanding the wave function in the driving frequency with 
$r_{\text{st}}(\mathcal{E}_l) = r_{\text{st}}(\mathcal{E}) +(\partial r_{\text{st}}/\partial \mathcal{E}) l \hbar \Omega$, we find
\begin{equation}
  \psi_r(t;\mathcal{E}) = a(t) r_{\text{st}}(\mathcal{E})
  + i \hbar \frac{\partial r_{\text{st}}}{\partial \mathcal{E}}  \frac{\partial a}{\partial t}.
  \label{eq:psi-r}
\end{equation}
The first term of Eq.~(\ref{eq:psi-r}) does not contribute to the mean number $\braket{n(\mathcal{E})}_{RL} +  \braket{n(\mathcal{E})}_{LL} $, as the contribution is proportional to $ \int_0^{2\pi/\Omega} dt \, \mathcal{V}_{\text{ac}}(t) |a(t)|^2 = \int_0^{2\pi/\Omega} dt \, \mathcal{V}_{\text{ac}}(t)= 0 $.
We now consider the second term,
\begin{align}
  \braket{n(\mathcal{E})}_{RL} +  \braket{n(\mathcal{E})}_{LL}
  &=-\frac{1}{2\pi}\int_0^{2\pi/\Omega} dt\,
    e \mathcal{V}_{\text{ac}}(t) \nonumber \\
  & \quad \times
    2\text{Re}\Big[ i r^*_{\text{st}}(\mathcal{E}) \frac{\partial r_{\text{st}}}{\partial \mathcal{E}} a^*(t) \frac{\partial a}{\partial t} \Big].
  \label{eq:nL-nonchi-4}
\end{align}
Using Eq.~(\ref{eq:d-ln-at}), we obtain 
\begin{equation}
  \braket{n(\mathcal{E})}_{LL}+\braket{n(\mathcal{E})}_{RL}
  =  -\frac{ \partial |r_{\text{st}}(\mathcal{E})|^2}{\partial \mathcal{E}} 
  \frac{e^2\overline{\mathcal{V}_{\text{ac}}^2(t)}}{\hbar \Omega}.
  \label{eq:nL-nonchi-slow}
\end{equation}
In the zero-temperature limit, $\theta_L \ll \Gamma$, we obtain a simple expression of the AC power injection, using Eq.~(\ref{eq:P-in}) and (\ref{eq:nL-nonchi-slow}),
\begin{equation}
  \overline{P_{\text{in}}}
  =  |t_{\text{st}}(\mu_L)|^2 \, \frac{e^2}{h}\, \overline{\mathcal{V}_{\text{ac}}^2(t)}
  \label{eq:Pin-nonchi-slow-0temp}
\end{equation}
This is the AC power determined by Joule's law with conductance $|t_{\text{st}}(\mu)|^2e^2/h$.
It shows that the AC power injection is generally nonzero for nonchiral setups, in contrast to the chiral setup.
One should not apply Eq.~(\ref{eq:Pin-nonchi-slow-0temp}) with $|t_{\text{st}}(\mu)|^2=1$ to the situation when there is no resonance level, because the Floquet scattering matrix becomes different, as
$\mathcal{S}_{RL}(\mathcal{E}_n, \mathcal{E}) = a_n $,
$\mathcal{S}_{LR}(\mathcal{E}_n, \mathcal{E}) = a_{-n}^*$,
$\mathcal{S}_{LL}(\mathcal{E}_n, \mathcal{E}) = \mathcal{S}_{RR}(\mathcal{E}_n, \mathcal{E})=0$,
which leads to $\overline{P_{\text{in}}}=0$.

\begin{figure}[t]
  \centering
  \includegraphics[width=\columnwidth]{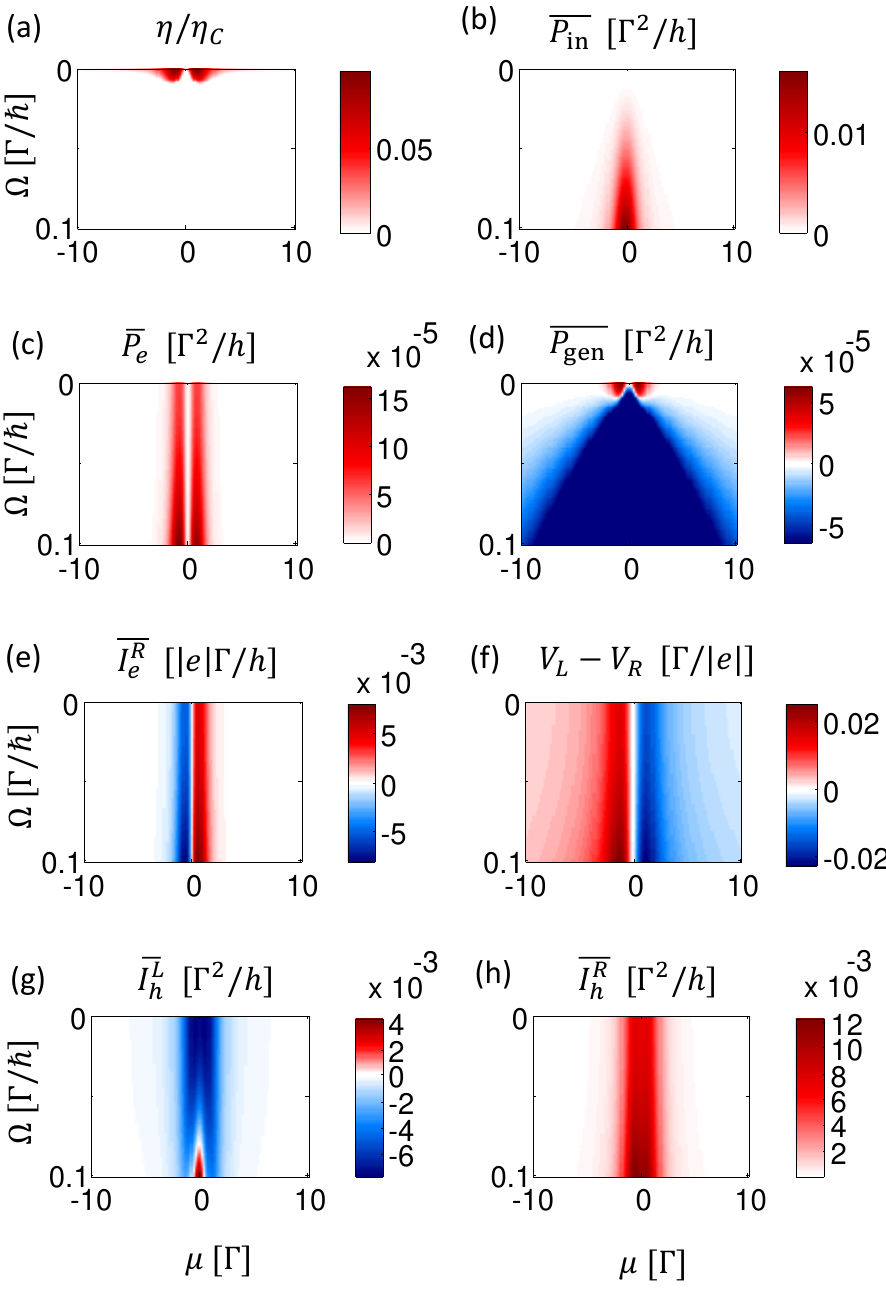}
  \caption{Heat engine efficiency in the AC driven nonchiral conductor of Supplementary Figure~\ref{fig:setup-nonchi}. (a) The efficiency decreases when the AC frequency increases, satisfying the Carnot limit, in contrast to the chiral-conductor. This is due to large positive AC power injection (b) which dominates the electric power (c) and makes the generated power (d) negative. Here, the generated power lower than $5\times 10^{-5} \Gamma^2/h$ is plotted with blue color. Additional details are also shown, such as charge current (e), DC voltage maximizing electric power chosen as in Eq.~(\ref{eq:dV-max-chiral}) (f), and heat current into the left (g), and right (h) reservoir. }
  \label{fig:eng_nonchi}
\end{figure}

Supplementary Figure~\ref{fig:eng_nonchi} shows the efficiency decrease by the AC driving in the case of nonchiral conductors. The  parameters such as AC driving protocol, temperatures $\theta_L$ and $\theta_R$ are the same as in Fig.~\ref{fig:eng-chiral}.

\section{Discussion on reservoirs with different chemical potentials}

Here we discuss another reservoir configuration, different from the one considered in the main text, but possible in experimental situations, where each reservoir can have arbitrary chemical potentials, see Supplementary Figure~\ref{fig:a}.
Below we show that any difference in the chemical potentials between the left two reservoirs diminishes the power generation as the potential bias between the two induces a power dissipation. This justifies our choice for the reservoir configuration in Fig.~\ref{fig:setup}(b).

\begin{figure}[h]
  \centering
  \includegraphics[width=\columnwidth]{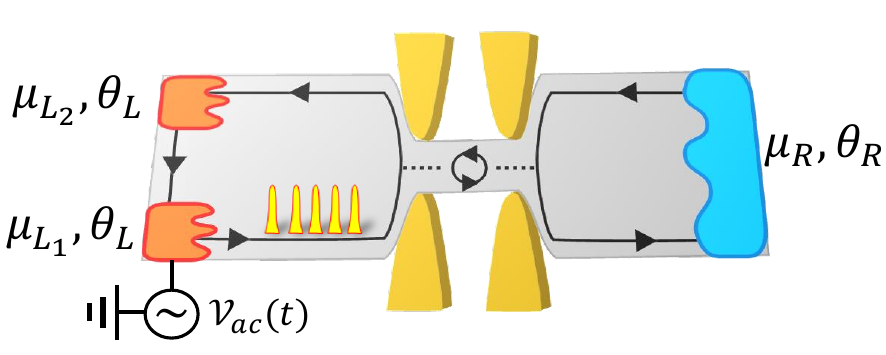}
  \caption{ Three terminal setup with arbitrary chemical potentials for each reservoir.}
  \label{fig:a}
\end{figure}

As in the main text, we employ the Floquet scattering formalism to calculate the currents into all the terminals.
The Floquet scattering matrix represented in the reservoir indexes ordered as $L_1$, $L_2$, and $R$ is
\begin{equation}
  \label{eq:S}
  \mathcal{S}_{\alpha \beta}(\mathcal{E}_n, \mathcal{E})
  =
  \begin{bmatrix}
    0 & a_{-n}^*  & 0  \\
     a_n r_{\text{st}}(\mathcal{E}_n) & 0 & \delta_{n0} t'_{\text{st}}(\mathcal{E}) \\
    a_n t_{\text{st}}(\mathcal{E}_n) & 0 & \delta_{n0} r'_{\text{st}}(\mathcal{E})
  \end{bmatrix} .
\end{equation}
The element $S_{L_1 L_2}(\mathcal{E}_n, \mathcal{E})$  is $a^*_{-n}$ because the electron enters into the region under the AC voltage (see Section PHOTOTRANSITION AMPLITUDE $a_n$ in SI).
This term does not contribute to the net heat and charge current, $\overline{I_h^L} = \overline{I_h^{L_1}}+\overline{I_h^{L_2}}$, $\overline{I_h^R}$, and $\overline{I_e^R}$.
We note that Eq.~(\ref{eq:S}) satisfies the unitarity conditions, $\sum_{n \alpha} |S_{\alpha \beta}(\mathcal{E}_n, \mathcal{E})|^2=1$ and $\sum_{n \beta} |S_{\alpha \beta}(\mathcal{E}, \mathcal{E}_{-n})|^2=1$.

The charge, heat, and energy currents into reservoir $\alpha$ are determined by Eq.~(\ref{eq:S}) and Eqs.~(\ref{eq:I-e-alpha})--~(\ref{eq:I-u-alpha}).
The power injection by AC voltage is obtained using the energy conservation relation, $\overline{P_{\text{in}}} = \sum_\alpha \overline{I_u^\alpha}$,
\begin{equation}
  \label{eq:4}
  \overline{P_{\text{in}}}
    = \int \frac{d \mathcal{E}}{h} \sum_{\alpha,\beta, n} n \hbar \Omega |S_{\alpha \beta}(\mathcal{E}_n, \mathcal{E}) |^2 f_\beta (\mathcal{E}) .
\end{equation}
This vanishes due to the chirality manifested in Eq.~(\ref{eq:S}), $\sum_n |a_n|^2 =1 $, and $\sum_n n |a_n|^2=0$,
\begin{equation}
  \label{eq:6}
  \overline{P_{\text{in}}}=0 .
\end{equation}

The electric power generated by the charge flow is
\begin{equation}
  \label{eq:5}
  \overline{P}_e = \sum_\alpha \mu_\alpha \overline{I_e^\alpha}/e .
\end{equation}

Using the form of the Floquet scattering matrix, Eq.~(\ref{eq:S}), we obtain the electric power,
\begin{equation}
  \label{eq:7}
  \begin{aligned}
  \overline{P_e}
  &= (\mu_R-\mu_{L_2}) \frac{\overline{I_e^R}}{e}  \\
  &\quad  + (\mu_{L_1}-\mu_{L_2}) \int \frac{ d \mathcal{E}}{h}
  \Big(- f_{L_1}(\mathcal{E}) + f_{L_2}(\mathcal{E}) \Big)  ,
  \end{aligned}
\end{equation}
The second term is a nonpositive quantity which describes the power dissipation between the reservoirs $L_1$ and $L_2$. It vanishes when the two reservoirs have the same chemical potential.

Therefore, the condition used in the main text, $\mu_{L_1}=\mu_{L_2}$, is the best choice for obtaining large electric power.
Defining $\mu_L \equiv \mu_{L_1}=\mu_{L_2}\equiv $ and $f_L(\mathcal{E}) \equiv f_{L_1}(\mathcal{E}) = f_{L_1}(\mathcal{E})$,
the electric power $\overline{P_e}$, the total heat current into the hot reservoir $\overline{I_h^L}\equiv \overline{I_h^{L_1}} +\overline{I_h^{L_2}}$, and the efficiency become equal to the quantities described in the main text.

\end{document}